\documentstyle{article}

\newcommand{\ptx}{{<}p_x^2{>}^{1/2}}
\newcommand{\ptxs}{{<}p_x^2{>}}
\newcommand{\pty}{{<}p_y^2{>}^{1/2}}
\newcommand{\ptys}{{<}p_y^2{>}}
\newcommand{\ptz}{{<}p_z^2{>}^{1/2}}
\newcommand{\ptzs}{{<}p_z^2{>}}
\newcommand{\dv}{\mathop{\rm div}\nolimits}
\newcommand{\kp}{\rule[-0.49pt]{0.98pt}{0.98pt}}

\begin{document}

\begin{center}
\LARGE
Possible asymmetry of particle distribution around axes of hard jets
in ultrarelativistic heavy ion collisions
\end{center}

\vspace{4.5cm}
\begin{center}
\large
M. Yu. Kopysov and Yu. E. Pokrovsky
\end{center}
\vspace{4.5cm}
\begin{center} \large
Russian Research Centre "Kurchatov Institute", \\
123182, Moscow, Russia
\end{center}
\newpage
\large

\begin{abstract}
We discuss a possible new manifestation of the formation of an initial
nonequilibrium flux tube stage in high energy heavy ion collisions. It
is shown that a strong asymmetry in particle distributions around axes
of hard-transverse jets takes place due to a large difference between
longitudinal and transverse forces acting on the hardly scattered
quarks or gluons crossing the flux tubes.
\end{abstract}

Jet formation and evolution in hot-dense hadron matter has been a
subject of recent papers. This problem is considered in many studies:
interaction of the leading partons of di-jets with the excited matter
produced in ultrarelativistic nucleus-nucleus collisions leads to
energy-momentum transfer to the hot and dense medium, which may
manifest itself as jet quenching \cite{que1,que2}, a broadening of the
acoplanarity distribution \cite{acop1,acop2,acop3} and possible
appearance of dynamic instabilities in quark-gluon plasma (QGP)
\cite{Silin,Selikhov,Pav} which leads to characteristic pion and photon
emission. Also the quantum--chromodynamics (QCD) flux tube dynamics at
finite temperature can lead to broader $p_T$ distribution and increased
strangeness yield \cite{6}.

\begin{figure}

\setlength{\unitlength}{0.1mm}  \begin{center}
\begin{picture}(850,600)(-86,0) \thicklines
\put(75,175){\oval(150,350)}    \put(575,175){\oval(150,350)}
\put(50,300){\circle*{30}}      \put(200,300){\circle{30}}
\put(240,300){\circle*{30}}     \put(560,300){\circle{30}}
\put(65,300){\line(1,0){120}}   \put(255,300){\line(1,0){290}}

\put(100,250){\circle{30}}      \put(380,250){\circle*{30}}
\put(420,250){\circle{30}}      \put(610,250){\circle*{30}}
\put(115,250){\line(1,0){250}}  \put(435,250){\line(1,0){160}}

\put(30,200){\circle{30}}       \put(180,200){\circle*{30}}
\put(220,200){\circle{30}}      \put(410,200){\circle*{30}}
\put(450,200){\circle{30}}      \put(590,200){\circle*{30}}
\put(45,200){\line(1,0){120}}   \put(235,200){\line(1,0){160}}
\put(465,200){\line(1,0){110}}

\put(40,150){\circle*{30}}      \put(380,150){\circle{30}}
\put(420,150){\circle*{30}}     \put(620,150){\circle{30}}
\put(55,150){\line(1,0){310}}   \put(435,150){\line(1,0){170}}

\put(90,100){\circle{30}}       \put(190,100){\circle*{30}}
\put(230,100){\circle{30}}      \put(340,100){\circle*{30}}
\put(380,100){\circle{30}}      \put(530,100){\circle*{30}}
\put(105,100){\line(1,0){70}}   \put(245,100){\line(1,0){80}}
\put(395,100){\line(1,0){120}}

\put(60,50){\circle*{30}}       \put(290,50){\circle{30}}
\put(330,50){\circle*{30}}      \put(590,50){\circle{30}}
\put(75,50){\line(1,0){200}}    \put(345,50){\line(1,0){230}}

\put(310,125){\vector(0,1){445}}\thinlines
\thicklines
\put(310,175){\line(-1,1){255}}
\put(310,175){\line(1,1){255}}  \thinlines
\multiput(310,175)(-40,100){3}{\line(-2,5){22}}
\multiput(310,175)(40,100){3}{\line(2,5){22}}
\put(-30,175){\vector(1,0){770}}\put(720,145){$z$}
\put(665,222){beam}             \put(665,187){axis}

\put(210,380){\vector(2,1){200}}\put(410,470){$x$}
\put(18,430){\vector(1,0){584}} \put(592,440){$z$}
\put(280,550){$y$}              \put(340,550){jet axis}
\put(-122,370){\line(2,1){240}} \put(-122,370){\line(1,0){624}}
\put(118,490){\line(1,0){624}}  \put(742,490){\line(-2,-1){240}}

\thicklines
\put(-95,555){\circle*{30}}     \put(-65,540){3~(quark)}
\put(-95,510){\circle{30}}      \put(-65,495){$\bar 3~$(
$\bar q,~qq)$}

\put( 560.00, 430.00){\kp}      \put( 560.00, 430.00){\kp}
\put(  60.00, 430.00){\kp}      \put(  60.00, 430.00){\kp}
\put( 560.91, 430.47){\kp}      \put( 559.03, 429.53){\kp}
\put(  60.97, 430.47){\kp}      \put(  59.09, 429.53){\kp}
\put( 561.76, 430.94){\kp}      \put( 557.99, 429.06){\kp}
\put(  62.01, 430.94){\kp}      \put(  58.24, 429.06){\kp}
\put( 562.55, 431.41){\kp}      \put( 556.90, 428.59){\kp}
\put(  63.10, 431.41){\kp}      \put(  57.45, 428.59){\kp}
\put( 563.27, 431.88){\kp}      \put( 555.74, 428.12){\kp}
\put(  64.26, 431.88){\kp}      \put(  56.73, 428.12){\kp}
\put( 563.94, 432.35){\kp}      \put( 554.52, 427.65){\kp}
\put(  65.48, 432.35){\kp}      \put(  56.06, 427.65){\kp}
\put( 564.54, 432.82){\kp}      \put( 553.24, 427.18){\kp}
\put(  66.76, 432.82){\kp}      \put(  55.46, 427.18){\kp}
\put( 565.07, 433.29){\kp}      \put( 551.91, 426.71){\kp}
\put(  68.09, 433.29){\kp}      \put(  54.93, 426.71){\kp}
\put( 565.55, 433.76){\kp}      \put( 550.51, 426.24){\kp}
\put(  69.49, 433.76){\kp}      \put(  54.45, 426.24){\kp}
\put( 565.96, 434.23){\kp}      \put( 549.05, 425.77){\kp}
\put(  70.95, 434.23){\kp}      \put(  54.04, 425.77){\kp}
\put( 566.31, 434.69){\kp}      \put( 547.54, 425.31){\kp}
\put(  72.46, 434.69){\kp}      \put(  53.69, 425.31){\kp}
\put( 566.59, 435.16){\kp}      \put( 545.96, 424.84){\kp}
\put(  74.04, 435.16){\kp}      \put(  53.41, 424.84){\kp}
\put( 566.81, 435.62){\kp}      \put( 544.33, 424.38){\kp}
\put(  75.67, 435.62){\kp}      \put(  53.19, 424.38){\kp}
\put( 566.97, 436.08){\kp}      \put( 542.64, 423.92){\kp}
\put(  77.36, 436.08){\kp}      \put(  53.03, 423.92){\kp}
\put( 567.07, 436.54){\kp}      \put( 540.89, 423.46){\kp}
\put(  79.11, 436.54){\kp}      \put(  52.93, 423.46){\kp}
\put( 567.10, 437.00){\kp}      \put( 539.09, 423.00){\kp}
\put(  80.91, 437.00){\kp}      \put(  52.90, 423.00){\kp}
\put( 567.07, 437.46){\kp}      \put( 537.22, 422.54){\kp}
\put(  82.78, 437.46){\kp}      \put(  52.93, 422.54){\kp}
\put( 566.97, 437.92){\kp}      \put( 535.31, 422.08){\kp}
\put(  84.69, 437.92){\kp}      \put(  53.03, 422.08){\kp}
\put( 566.81, 438.37){\kp}      \put( 533.33, 421.63){\kp}
\put(  86.67, 438.37){\kp}      \put(  53.19, 421.63){\kp}
\put( 566.59, 438.82){\kp}      \put( 531.31, 421.18){\kp}
\put(  88.69, 438.82){\kp}      \put(  53.41, 421.18){\kp}
\put( 566.31, 439.27){\kp}      \put( 529.22, 420.73){\kp}
\put(  90.78, 439.27){\kp}      \put(  53.69, 420.73){\kp}
\put( 565.96, 439.72){\kp}      \put( 527.09, 420.28){\kp}
\put(  92.91, 439.72){\kp}      \put(  54.04, 420.28){\kp}
\put( 565.54, 440.16){\kp}      \put( 524.90, 419.84){\kp}
\put(  95.10, 440.16){\kp}      \put(  54.46, 419.84){\kp}
\put( 565.07, 440.60){\kp}      \put( 522.65, 419.40){\kp}
\put(  97.35, 440.60){\kp}      \put(  54.93, 419.40){\kp}
\put( 564.53, 441.04){\kp}      \put( 520.36, 418.96){\kp}
\put(  99.64, 441.04){\kp}      \put(  55.47, 418.96){\kp}
\put( 563.93, 441.48){\kp}      \put( 518.01, 418.52){\kp}
\put( 101.99, 441.48){\kp}      \put(  56.07, 418.52){\kp}
\put( 563.27, 441.91){\kp}      \put( 515.61, 418.09){\kp}
\put( 104.39, 441.91){\kp}      \put(  56.73, 418.09){\kp}
\put( 562.54, 442.35){\kp}      \put( 513.16, 417.65){\kp}
\put( 106.84, 442.35){\kp}      \put(  57.46, 417.65){\kp}
\put( 561.75, 442.77){\kp}      \put( 510.66, 417.23){\kp}
\put( 109.34, 442.77){\kp}      \put(  58.25, 417.23){\kp}
\put( 560.90, 443.20){\kp}      \put( 508.11, 416.80){\kp}
\put( 111.89, 443.20){\kp}      \put(  59.10, 416.80){\kp}
\put( 559.99, 443.62){\kp}      \put( 505.51, 416.38){\kp}
\put( 114.49, 443.62){\kp}      \put(  60.01, 416.38){\kp}
\put( 559.02, 444.04){\kp}      \put( 502.87, 415.96){\kp}
\put( 117.13, 444.04){\kp}      \put(  60.98, 415.96){\kp}
\put( 557.98, 444.45){\kp}      \put( 500.17, 415.55){\kp}
\put( 119.83, 444.45){\kp}      \put(  62.02, 415.55){\kp}
\put( 556.89, 444.86){\kp}      \put( 497.43, 415.14){\kp}
\put( 122.57, 444.86){\kp}      \put(  63.11, 415.14){\kp}
\put( 555.73, 445.27){\kp}      \put( 494.64, 414.73){\kp}
\put( 125.36, 445.27){\kp}      \put(  64.27, 414.73){\kp}
\put( 554.51, 445.67){\kp}      \put( 491.81, 414.33){\kp}
\put( 128.19, 445.67){\kp}      \put(  65.49, 414.33){\kp}
\put( 553.23, 446.07){\kp}      \put( 488.93, 413.93){\kp}
\put( 131.07, 446.07){\kp}      \put(  66.77, 413.93){\kp}
\put( 551.89, 446.47){\kp}      \put( 486.01, 413.53){\kp}
\put( 133.99, 446.47){\kp}      \put(  68.11, 413.53){\kp}
\put( 550.50, 446.86){\kp}      \put( 483.05, 413.14){\kp}
\put( 136.95, 446.86){\kp}      \put(  69.50, 413.14){\kp}
\put( 549.04, 447.25){\kp}      \put( 480.04, 412.75){\kp}
\put( 139.96, 447.25){\kp}      \put(  70.96, 412.75){\kp}
\put( 547.52, 447.63){\kp}      \put( 476.99, 412.37){\kp}
\put( 143.01, 447.63){\kp}      \put(  72.48, 412.37){\kp}
\put( 545.95, 448.01){\kp}      \put( 473.90, 411.99){\kp}
\put( 146.10, 448.01){\kp}      \put(  74.05, 411.99){\kp}
\put( 544.31, 448.39){\kp}      \put( 470.76, 411.61){\kp}
\put( 149.24, 448.39){\kp}      \put(  75.69, 411.61){\kp}
\put( 542.62, 448.76){\kp}      \put( 467.59, 411.24){\kp}
\put( 152.41, 448.76){\kp}      \put(  77.38, 411.24){\kp}
\put( 540.87, 449.12){\kp}      \put( 464.38, 410.88){\kp}
\put( 155.62, 449.12){\kp}      \put(  79.13, 410.88){\kp}
\put( 539.07, 449.48){\kp}      \put( 461.13, 410.52){\kp}
\put( 158.87, 449.48){\kp}      \put(  80.93, 410.52){\kp}
\put( 537.21, 449.84){\kp}      \put( 457.85, 410.16){\kp}
\put( 162.15, 449.84){\kp}      \put(  82.79, 410.16){\kp}
\put( 535.29, 450.19){\kp}      \put( 454.53, 409.81){\kp}
\put( 165.47, 450.19){\kp}      \put(  84.71, 409.81){\kp}
\put( 533.32, 450.54){\kp}      \put( 451.17, 409.46){\kp}
\put( 168.83, 450.54){\kp}      \put(  86.69, 409.46){\kp}
\put( 531.29, 450.88){\kp}      \put( 447.78, 409.12){\kp}
\put( 172.22, 450.88){\kp}      \put(  88.71, 409.12){\kp}
\put( 529.20, 451.21){\kp}      \put( 444.35, 408.79){\kp}
\put( 175.65, 451.21){\kp}      \put(  90.80, 408.79){\kp}
\put( 527.07, 451.54){\kp}      \put( 440.89, 408.46){\kp}
\put( 179.11, 451.54){\kp}      \put(  92.93, 408.46){\kp}
\put( 524.87, 451.87){\kp}      \put( 437.40, 408.13){\kp}
\put( 182.60, 451.87){\kp}      \put(  95.13, 408.13){\kp}
\put( 522.63, 452.19){\kp}      \put( 433.88, 407.81){\kp}
\put( 186.12, 452.19){\kp}      \put(  97.37, 407.81){\kp}
\put( 520.33, 452.50){\kp}      \put( 430.32, 407.50){\kp}
\put( 189.68, 452.50){\kp}      \put(  99.67, 407.50){\kp}
\put( 517.99, 452.81){\kp}      \put( 426.74, 407.19){\kp}
\put( 193.26, 452.81){\kp}      \put( 102.01, 407.19){\kp}
\put( 515.59, 453.12){\kp}      \put( 423.13, 406.88){\kp}
\put( 196.87, 453.12){\kp}      \put( 104.41, 406.88){\kp}
\put( 513.14, 453.41){\kp}      \put( 419.48, 406.59){\kp}
\put( 200.52, 453.41){\kp}      \put( 106.86, 406.59){\kp}
\put( 510.64, 453.70){\kp}      \put( 415.82, 406.30){\kp}
\put( 204.18, 453.70){\kp}      \put( 109.36, 406.30){\kp}
\put( 508.09, 453.99){\kp}      \put( 412.12, 406.01){\kp}
\put( 207.88, 453.99){\kp}      \put( 111.91, 406.01){\kp}
\put( 505.49, 454.27){\kp}      \put( 408.41, 405.73){\kp}
\put( 211.59, 454.27){\kp}      \put( 114.51, 405.73){\kp}
\put( 502.84, 454.54){\kp}      \put( 404.66, 405.46){\kp}
\put( 215.34, 454.54){\kp}      \put( 117.16, 405.46){\kp}
\put( 500.15, 454.81){\kp}      \put( 400.90, 405.19){\kp}
\put( 219.10, 454.81){\kp}      \put( 119.85, 405.19){\kp}
\put( 497.40, 455.07){\kp}      \put( 397.11, 404.93){\kp}
\put( 222.89, 455.07){\kp}      \put( 122.60, 404.93){\kp}
\put( 494.62, 455.33){\kp}      \put( 393.30, 404.67){\kp}
\put( 226.70, 455.33){\kp}      \put( 125.38, 404.67){\kp}
\put( 491.78, 455.58){\kp}      \put( 389.47, 404.42){\kp}
\put( 230.53, 455.58){\kp}      \put( 128.22, 404.42){\kp}
\put( 488.90, 455.82){\kp}      \put( 385.62, 404.18){\kp}
\put( 234.38, 455.82){\kp}      \put( 131.10, 404.18){\kp}
\put( 485.98, 456.06){\kp}      \put( 381.75, 403.94){\kp}
\put( 238.25, 456.06){\kp}      \put( 134.02, 403.94){\kp}
\put( 483.02, 456.29){\kp}      \put( 377.86, 403.71){\kp}
\put( 242.14, 456.29){\kp}      \put( 136.98, 403.71){\kp}
\put( 480.01, 456.51){\kp}      \put( 373.96, 403.49){\kp}
\put( 246.04, 456.51){\kp}      \put( 139.99, 403.49){\kp}
\put( 476.96, 456.73){\kp}      \put( 370.04, 403.27){\kp}
\put( 249.96, 456.73){\kp}      \put( 143.04, 403.27){\kp}
\put( 473.87, 456.94){\kp}      \put( 366.10, 403.06){\kp}
\put( 253.90, 456.94){\kp}      \put( 146.13, 403.06){\kp}
\put( 470.73, 457.14){\kp}      \put( 362.16, 402.86){\kp}
\put( 257.84, 457.14){\kp}      \put( 149.27, 402.86){\kp}
\put( 467.56, 457.34){\kp}      \put( 358.19, 402.66){\kp}
\put( 261.81, 457.34){\kp}      \put( 152.44, 402.66){\kp}
\put( 464.35, 457.53){\kp}      \put( 354.22, 402.47){\kp}
\put( 265.78, 457.53){\kp}      \put( 155.65, 402.47){\kp}
\put( 461.10, 457.72){\kp}      \put( 350.24, 402.28){\kp}
\put( 269.76, 457.72){\kp}      \put( 158.90, 402.28){\kp}
\put( 457.82, 457.89){\kp}      \put( 346.24, 402.11){\kp}
\put( 273.76, 457.89){\kp}      \put( 162.18, 402.11){\kp}
\put( 454.50, 458.06){\kp}      \put( 342.24, 401.94){\kp}
\put( 277.76, 458.06){\kp}      \put( 165.50, 401.94){\kp}
\put( 451.14, 458.23){\kp}      \put( 338.23, 401.77){\kp}
\put( 281.77, 458.23){\kp}      \put( 168.86, 401.77){\kp}
\put( 447.74, 458.38){\kp}      \put( 334.21, 401.62){\kp}
\put( 285.79, 458.38){\kp}      \put( 172.26, 401.62){\kp}
\put( 444.32, 458.53){\kp}      \put( 330.19, 401.47){\kp}
\put( 289.81, 458.53){\kp}      \put( 175.68, 401.47){\kp}
\put( 440.86, 458.67){\kp}      \put( 326.16, 401.33){\kp}
\put( 293.84, 458.67){\kp}      \put( 179.14, 401.33){\kp}
\put( 437.37, 458.81){\kp}      \put( 322.13, 401.19){\kp}
\put( 297.87, 458.81){\kp}      \put( 182.63, 401.19){\kp}
\put( 433.84, 458.94){\kp}      \put( 318.09, 401.06){\kp}
\put( 301.91, 458.94){\kp}      \put( 186.16, 401.06){\kp}
\put( 430.29, 459.06){\kp}      \put( 314.06, 400.94){\kp}
\put( 305.94, 459.06){\kp}      \put( 189.71, 400.94){\kp}
\put( 426.70, 459.17){\kp}      \put( 310.02, 400.83){\kp}
\put( 309.98, 459.17){\kp}      \put( 193.30, 400.83){\kp}
\put( 423.09, 459.28){\kp}      \put( 305.98, 400.72){\kp}
\put( 314.02, 459.28){\kp}      \put( 196.91, 400.72){\kp}
\put( 419.45, 459.38){\kp}      \put( 301.94, 400.62){\kp}
\put( 318.06, 459.38){\kp}      \put( 200.55, 400.62){\kp}
\put( 415.78, 459.47){\kp}      \put( 297.91, 400.53){\kp}
\put( 322.09, 459.47){\kp}      \put( 204.22, 400.53){\kp}
\put( 412.09, 459.55){\kp}      \put( 293.88, 400.45){\kp}
\put( 326.12, 459.55){\kp}      \put( 207.91, 400.45){\kp}
\put( 408.37, 459.63){\kp}      \put( 289.85, 400.37){\kp}
\put( 330.15, 459.63){\kp}      \put( 211.63, 400.37){\kp}
\put( 404.63, 459.70){\kp}      \put( 285.82, 400.30){\kp}
\put( 334.18, 459.70){\kp}      \put( 215.37, 400.30){\kp}
\put( 400.86, 459.76){\kp}      \put( 281.81, 400.24){\kp}
\put( 338.19, 459.76){\kp}      \put( 219.14, 400.24){\kp}
\put( 397.07, 459.82){\kp}      \put( 277.80, 400.18){\kp}
\put( 342.20, 459.82){\kp}      \put( 222.93, 400.18){\kp}
\put( 393.26, 459.87){\kp}      \put( 273.79, 400.13){\kp}
\put( 346.21, 459.87){\kp}      \put( 226.74, 400.13){\kp}
\put( 389.43, 459.91){\kp}      \put( 269.80, 400.09){\kp}
\put( 350.20, 459.91){\kp}      \put( 230.57, 400.09){\kp}
\put( 385.58, 459.94){\kp}      \put( 265.82, 400.06){\kp}
\put( 354.18, 459.94){\kp}      \put( 234.42, 400.06){\kp}
\put( 381.71, 459.97){\kp}      \put( 261.84, 400.03){\kp}
\put( 358.16, 459.97){\kp}      \put( 238.29, 400.03){\kp}
\put( 377.82, 459.99){\kp}      \put( 257.88, 400.01){\kp}
\put( 362.12, 459.99){\kp}      \put( 242.18, 400.01){\kp}
\put( 373.92, 460.00){\kp}      \put( 253.93, 400.00){\kp}
\put( 366.07, 460.00){\kp}      \put( 246.08, 400.00){\kp}
\put( 370.00, 460.00){\kp}      \put( 250.00, 400.00){\kp}
\put( 370.00, 460.00){\kp}      \put( 250.00, 400.00){\kp}
\renewcommand{\kp}{\rule[-0.225pt]{0.45pt}{0.45pt}}
\put( 410.00, 430.00){\kp}      \put( 410.00, 430.00){\kp}
\put( 210.00, 430.00){\kp}      \put( 210.00, 430.00){\kp}
\put( 410.93, 430.47){\kp}      \put( 409.05, 429.53){\kp}
\put( 210.95, 430.47){\kp}      \put( 209.07, 429.53){\kp}
\put( 411.84, 430.94){\kp}      \put( 408.07, 429.06){\kp}
\put( 211.93, 430.94){\kp}      \put( 208.16, 429.06){\kp}
\put( 412.72, 431.41){\kp}      \put( 407.06, 428.59){\kp}
\put( 212.94, 431.41){\kp}      \put( 207.28, 428.59){\kp}
\put( 413.57, 431.88){\kp}      \put( 406.04, 428.12){\kp}
\put( 213.96, 431.88){\kp}      \put( 206.43, 428.12){\kp}
\put( 414.40, 432.35){\kp}      \put( 404.98, 427.65){\kp}
\put( 215.02, 432.35){\kp}      \put( 205.60, 427.65){\kp}
\put( 415.20, 432.82){\kp}      \put( 403.91, 427.18){\kp}
\put( 216.09, 432.82){\kp}      \put( 204.80, 427.18){\kp}
\put( 415.98, 433.29){\kp}      \put( 402.81, 426.71){\kp}
\put( 217.19, 433.29){\kp}      \put( 204.02, 426.71){\kp}
\put( 416.73, 433.76){\kp}      \put( 401.69, 426.24){\kp}
\put( 218.31, 433.76){\kp}      \put( 203.27, 426.24){\kp}
\put( 417.46, 434.23){\kp}      \put( 400.55, 425.77){\kp}
\put( 219.45, 434.23){\kp}      \put( 202.54, 425.77){\kp}
\put( 418.15, 434.69){\kp}      \put( 399.38, 425.31){\kp}
\put( 220.62, 434.69){\kp}      \put( 201.85, 425.31){\kp}
\put( 418.83, 435.16){\kp}      \put( 398.20, 424.84){\kp}
\put( 221.80, 435.16){\kp}      \put( 201.17, 424.84){\kp}
\put( 419.47, 435.62){\kp}      \put( 396.99, 424.38){\kp}
\put( 223.01, 435.62){\kp}      \put( 200.53, 424.38){\kp}
\put( 420.09, 436.08){\kp}      \put( 395.76, 423.92){\kp}
\put( 224.24, 436.08){\kp}      \put( 199.91, 423.92){\kp}
\put( 420.68, 436.54){\kp}      \put( 394.50, 423.46){\kp}
\put( 225.50, 436.54){\kp}      \put( 199.32, 423.46){\kp}
\put( 421.24, 437.00){\kp}      \put( 393.23, 423.00){\kp}
\put( 226.77, 437.00){\kp}      \put( 198.76, 423.00){\kp}
\put( 421.78, 437.46){\kp}      \put( 391.94, 422.54){\kp}
\put( 228.06, 437.46){\kp}      \put( 198.22, 422.54){\kp}
\put( 422.29, 437.92){\kp}      \put( 390.62, 422.08){\kp}
\put( 229.38, 437.92){\kp}      \put( 197.71, 422.08){\kp}
\put( 422.77, 438.37){\kp}      \put( 389.29, 421.63){\kp}
\put( 230.71, 438.37){\kp}      \put( 197.23, 421.63){\kp}
\put( 423.22, 438.82){\kp}      \put( 387.94, 421.18){\kp}
\put( 232.06, 438.82){\kp}      \put( 196.78, 421.18){\kp}
\put( 423.65, 439.27){\kp}      \put( 386.56, 420.73){\kp}
\put( 233.44, 439.27){\kp}      \put( 196.35, 420.73){\kp}
\put( 424.04, 439.72){\kp}      \put( 385.17, 420.28){\kp}
\put( 234.83, 439.72){\kp}      \put( 195.96, 420.28){\kp}
\put( 424.41, 440.16){\kp}      \put( 383.76, 419.84){\kp}
\put( 236.24, 440.16){\kp}      \put( 195.59, 419.84){\kp}
\put( 424.75, 440.60){\kp}      \put( 382.34, 419.40){\kp}
\put( 237.66, 440.60){\kp}      \put( 195.25, 419.40){\kp}
\put( 425.07, 441.04){\kp}      \put( 380.89, 418.96){\kp}
\put( 239.11, 441.04){\kp}      \put( 194.93, 418.96){\kp}
\put( 425.35, 441.48){\kp}      \put( 379.43, 418.52){\kp}
\put( 240.57, 441.48){\kp}      \put( 194.65, 418.52){\kp}
\put( 425.60, 441.91){\kp}      \put( 377.95, 418.09){\kp}
\put( 242.05, 441.91){\kp}      \put( 194.40, 418.09){\kp}
\put( 425.83, 442.35){\kp}      \put( 376.45, 417.65){\kp}
\put( 243.55, 442.35){\kp}      \put( 194.17, 417.65){\kp}
\put( 426.03, 442.77){\kp}      \put( 374.94, 417.23){\kp}
\put( 245.06, 442.77){\kp}      \put( 193.97, 417.23){\kp}
\put( 426.20, 443.20){\kp}      \put( 373.41, 416.80){\kp}
\put( 246.59, 443.20){\kp}      \put( 193.80, 416.80){\kp}
\put( 426.34, 443.62){\kp}      \put( 371.86, 416.38){\kp}
\put( 248.14, 443.62){\kp}      \put( 193.66, 416.38){\kp}
\put( 426.45, 444.04){\kp}      \put( 370.30, 415.96){\kp}
\put( 249.70, 444.04){\kp}      \put( 193.55, 415.96){\kp}
\put( 426.54, 444.45){\kp}      \put( 368.73, 415.55){\kp}
\put( 251.27, 444.45){\kp}      \put( 193.46, 415.55){\kp}
\put( 426.59, 444.86){\kp}      \put( 367.14, 415.14){\kp}
\put( 252.86, 444.86){\kp}      \put( 193.41, 415.14){\kp}
\put( 426.62, 445.27){\kp}      \put( 365.53, 414.73){\kp}
\put( 254.47, 445.27){\kp}      \put( 193.38, 414.73){\kp}
\put( 426.61, 445.67){\kp}      \put( 363.91, 414.33){\kp}
\put( 256.09, 445.67){\kp}      \put( 193.39, 414.33){\kp}
\put( 426.58, 446.07){\kp}      \put( 362.28, 413.93){\kp}
\put( 257.72, 446.07){\kp}      \put( 193.42, 413.93){\kp}
\put( 426.52, 446.47){\kp}      \put( 360.64, 413.53){\kp}
\put( 259.36, 446.47){\kp}      \put( 193.48, 413.53){\kp}
\put( 426.43, 446.86){\kp}      \put( 358.98, 413.14){\kp}
\put( 261.02, 446.86){\kp}      \put( 193.57, 413.14){\kp}
\put( 426.32, 447.25){\kp}      \put( 357.31, 412.75){\kp}
\put( 262.69, 447.25){\kp}      \put( 193.68, 412.75){\kp}
\put( 426.17, 447.63){\kp}      \put( 355.63, 412.37){\kp}
\put( 264.37, 447.63){\kp}      \put( 193.83, 412.37){\kp}
\put( 425.99, 448.01){\kp}      \put( 353.94, 411.99){\kp}
\put( 266.06, 448.01){\kp}      \put( 194.01, 411.99){\kp}
\put( 425.79, 448.39){\kp}      \put( 352.24, 411.61){\kp}
\put( 267.76, 448.39){\kp}      \put( 194.21, 411.61){\kp}
\put( 425.56, 448.76){\kp}      \put( 350.53, 411.24){\kp}
\put( 269.47, 448.76){\kp}      \put( 194.44, 411.24){\kp}
\put( 425.30, 449.12){\kp}      \put( 348.81, 410.88){\kp}
\put( 271.19, 449.12){\kp}      \put( 194.70, 410.88){\kp}
\put( 425.01, 449.48){\kp}      \put( 347.07, 410.52){\kp}
\put( 272.93, 449.48){\kp}      \put( 194.99, 410.52){\kp}
\put( 424.69, 449.84){\kp}      \put( 345.33, 410.16){\kp}
\put( 274.67, 449.84){\kp}      \put( 195.31, 410.16){\kp}
\put( 424.34, 450.19){\kp}      \put( 343.58, 409.81){\kp}
\put( 276.42, 450.19){\kp}      \put( 195.66, 409.81){\kp}
\put( 423.97, 450.54){\kp}      \put( 341.82, 409.46){\kp}
\put( 278.18, 450.54){\kp}      \put( 196.03, 409.46){\kp}
\put( 423.57, 450.88){\kp}      \put( 340.06, 409.12){\kp}
\put( 279.94, 450.88){\kp}      \put( 196.43, 409.12){\kp}
\put( 423.14, 451.21){\kp}      \put( 338.28, 408.79){\kp}
\put( 281.72, 451.21){\kp}      \put( 196.86, 408.79){\kp}
\put( 422.68, 451.54){\kp}      \put( 336.50, 408.46){\kp}
\put( 283.50, 451.54){\kp}      \put( 197.32, 408.46){\kp}
\put( 422.19, 451.87){\kp}      \put( 334.72, 408.13){\kp}
\put( 285.28, 451.87){\kp}      \put( 197.81, 408.13){\kp}
\put( 421.68, 452.19){\kp}      \put( 332.92, 407.81){\kp}
\put( 287.08, 452.19){\kp}      \put( 198.32, 407.81){\kp}
\put( 421.14, 452.50){\kp}      \put( 331.12, 407.50){\kp}
\put( 288.88, 452.50){\kp}      \put( 198.86, 407.50){\kp}
\put( 420.57, 452.81){\kp}      \put( 329.32, 407.19){\kp}
\put( 290.68, 452.81){\kp}      \put( 199.43, 407.19){\kp}
\put( 419.97, 453.12){\kp}      \put( 327.51, 406.88){\kp}
\put( 292.49, 453.12){\kp}      \put( 200.03, 406.88){\kp}
\put( 419.35, 453.41){\kp}      \put( 325.70, 406.59){\kp}
\put( 294.30, 453.41){\kp}      \put( 200.65, 406.59){\kp}
\put( 418.70, 453.70){\kp}      \put( 323.88, 406.30){\kp}
\put( 296.12, 453.70){\kp}      \put( 201.30, 406.30){\kp}
\put( 418.02, 453.99){\kp}      \put( 322.06, 406.01){\kp}
\put( 297.94, 453.99){\kp}      \put( 201.98, 406.01){\kp}
\put( 417.32, 454.27){\kp}      \put( 320.24, 405.73){\kp}
\put( 299.76, 454.27){\kp}      \put( 202.68, 405.73){\kp}
\put( 416.59, 454.54){\kp}      \put( 318.41, 405.46){\kp}
\put( 301.59, 454.54){\kp}      \put( 203.41, 405.46){\kp}
\put( 415.83, 454.81){\kp}      \put( 316.58, 405.19){\kp}
\put( 303.42, 454.81){\kp}      \put( 204.17, 405.19){\kp}
\put( 415.05, 455.07){\kp}      \put( 314.75, 404.93){\kp}
\put( 305.25, 455.07){\kp}      \put( 204.95, 404.93){\kp}
\put( 414.24, 455.33){\kp}      \put( 312.92, 404.67){\kp}
\put( 307.08, 455.33){\kp}      \put( 205.76, 404.67){\kp}
\put( 413.41, 455.58){\kp}      \put( 311.09, 404.42){\kp}
\put( 308.91, 455.58){\kp}      \put( 206.59, 404.42){\kp}
\put( 412.55, 455.82){\kp}      \put( 309.26, 404.18){\kp}
\put( 310.74, 455.82){\kp}      \put( 207.45, 404.18){\kp}
\put( 411.66, 456.06){\kp}      \put( 307.43, 403.94){\kp}
\put( 312.57, 456.06){\kp}      \put( 208.34, 403.94){\kp}
\put( 410.75, 456.29){\kp}      \put( 305.60, 403.71){\kp}
\put( 314.40, 456.29){\kp}      \put( 209.25, 403.71){\kp}
\put( 409.82, 456.51){\kp}      \put( 303.77, 403.49){\kp}
\put( 316.23, 456.51){\kp}      \put( 210.18, 403.49){\kp}
\put( 408.86, 456.73){\kp}      \put( 301.94, 403.27){\kp}
\put( 318.06, 456.73){\kp}      \put( 211.14, 403.27){\kp}
\put( 407.88, 456.94){\kp}      \put( 300.11, 403.06){\kp}
\put( 319.89, 456.94){\kp}      \put( 212.12, 403.06){\kp}
\put( 406.87, 457.14){\kp}      \put( 298.29, 402.86){\kp}
\put( 321.71, 457.14){\kp}      \put( 213.13, 402.86){\kp}
\put( 405.84, 457.34){\kp}      \put( 296.47, 402.66){\kp}
\put( 323.53, 457.34){\kp}      \put( 214.16, 402.66){\kp}
\put( 404.78, 457.53){\kp}      \put( 294.65, 402.47){\kp}
\put( 325.35, 457.53){\kp}      \put( 215.22, 402.47){\kp}
\put( 403.70, 457.72){\kp}      \put( 292.84, 402.28){\kp}
\put( 327.16, 457.72){\kp}      \put( 216.30, 402.28){\kp}
\put( 402.60, 457.89){\kp}      \put( 291.03, 402.11){\kp}
\put( 328.97, 457.89){\kp}      \put( 217.40, 402.11){\kp}
\put( 401.47, 458.06){\kp}      \put( 289.22, 401.94){\kp}
\put( 330.78, 458.06){\kp}      \put( 218.53, 401.94){\kp}
\put( 400.33, 458.23){\kp}      \put( 287.42, 401.77){\kp}
\put( 332.58, 458.23){\kp}      \put( 219.67, 401.77){\kp}
\put( 399.16, 458.38){\kp}      \put( 285.63, 401.62){\kp}
\put( 334.37, 458.38){\kp}      \put( 220.84, 401.62){\kp}
\put( 397.97, 458.53){\kp}      \put( 283.84, 401.47){\kp}
\put( 336.16, 458.53){\kp}      \put( 222.03, 401.47){\kp}
\put( 396.75, 458.67){\kp}      \put( 282.06, 401.33){\kp}
\put( 337.94, 458.67){\kp}      \put( 223.25, 401.33){\kp}
\put( 395.52, 458.81){\kp}      \put( 280.28, 401.19){\kp}
\put( 339.72, 458.81){\kp}      \put( 224.48, 401.19){\kp}
\put( 394.26, 458.94){\kp}      \put( 278.51, 401.06){\kp}
\put( 341.49, 458.94){\kp}      \put( 225.74, 401.06){\kp}
\put( 392.98, 459.06){\kp}      \put( 276.75, 400.94){\kp}
\put( 343.25, 459.06){\kp}      \put( 227.02, 400.94){\kp}
\put( 391.69, 459.17){\kp}      \put( 275.00, 400.83){\kp}
\put( 345.00, 459.17){\kp}      \put( 228.31, 400.83){\kp}
\put( 390.37, 459.28){\kp}      \put( 273.26, 400.72){\kp}
\put( 346.74, 459.28){\kp}      \put( 229.63, 400.72){\kp}
\put( 389.03, 459.38){\kp}      \put( 271.53, 400.62){\kp}
\put( 348.47, 459.38){\kp}      \put( 230.97, 400.62){\kp}
\put( 387.68, 459.47){\kp}      \put( 269.80, 400.53){\kp}
\put( 350.20, 459.47){\kp}      \put( 232.32, 400.53){\kp}
\put( 386.30, 459.55){\kp}      \put( 268.09, 400.45){\kp}
\put( 351.91, 459.55){\kp}      \put( 233.70, 400.45){\kp}
\put( 384.90, 459.63){\kp}      \put( 266.38, 400.37){\kp}
\put( 353.62, 459.63){\kp}      \put( 235.10, 400.37){\kp}
\put( 383.49, 459.70){\kp}      \put( 264.69, 400.30){\kp}
\put( 355.31, 459.70){\kp}      \put( 236.51, 400.30){\kp}
\put( 382.06, 459.76){\kp}      \put( 263.01, 400.24){\kp}
\put( 356.99, 459.76){\kp}      \put( 237.94, 400.24){\kp}
\put( 380.61, 459.82){\kp}      \put( 261.34, 400.18){\kp}
\put( 358.66, 459.82){\kp}      \put( 239.39, 400.18){\kp}
\put( 379.14, 459.87){\kp}      \put( 259.68, 400.13){\kp}
\put( 360.32, 459.87){\kp}      \put( 240.86, 400.13){\kp}
\put( 377.66, 459.91){\kp}      \put( 258.03, 400.09){\kp}
\put( 361.97, 459.91){\kp}      \put( 242.34, 400.09){\kp}
\put( 376.16, 459.94){\kp}      \put( 256.40, 400.06){\kp}
\put( 363.60, 459.94){\kp}      \put( 243.84, 400.06){\kp}
\put( 374.64, 459.97){\kp}      \put( 254.78, 400.03){\kp}
\put( 365.22, 459.97){\kp}      \put( 245.36, 400.03){\kp}
\put( 373.11, 459.99){\kp}      \put( 253.17, 400.01){\kp}
\put( 366.83, 459.99){\kp}      \put( 246.89, 400.01){\kp}
\put( 371.56, 460.00){\kp}      \put( 251.58, 400.00){\kp}
\put( 368.42, 460.00){\kp}      \put( 248.44, 400.00){\kp}
\put( 370.00, 460.00){\kp}      \put( 250.00, 400.00){\kp}
\put( 370.00, 460.00){\kp}      \put( 250.00, 400.00){\kp}
\end{picture} \end{center}
\caption{
An illustration for the asymmetry of the secondary particle
distribution in the hard jets and the initial nonequilibrium flux tube
stage. The cone of dashed lines corresponds to the jet in the case of
no flux tube stage and the cone of full lines corresponds to the
asymmetric jet. Black circles denote quarks (triplet color
representation) and white circles denote antiquarks or diquarks
(antitriplet representation).
\label{fig1}}
\end{figure}
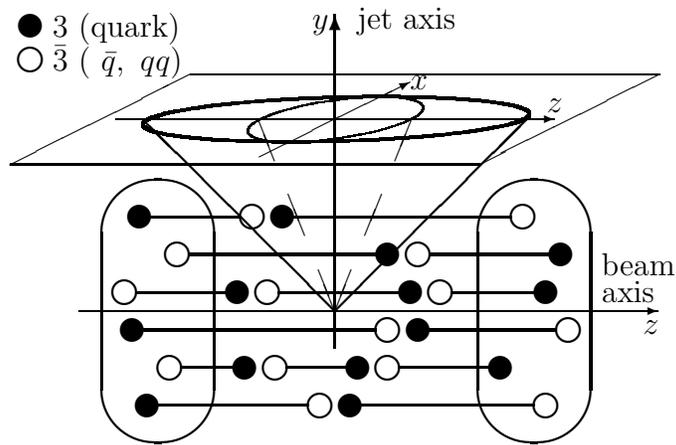

In this paper we suggest a new signature of the formation of an initial
flux tube like stage \cite{7}. It is shown that presence of the flux
tube stage in collisions leads to a modification of distributions of
secondary particles in hard transverse jets. In contrast to a single
jet case, in which the distributions are axially symmetrical relative
to the jet axis, the modification appears as the symmetry violation
(Fig.~\ref{fig1}). This phenomenon should take place due to the large
difference between the longitudinal and transverse forces acting on the
hardly scattered and almost transversely moved quarks or gluons which
crossed the flux tubes and formed the jets. So QCD hard jets can
serve as probes of the hot and dense matter in the flux tube stage.

Let us assume that formation of the multi flux tube state in
nucleus--nucleus collisions takes place at the initial stage: during
the collision of two heavy nuclei a system of flux tubes begin to
form in the volume of the overlap of the colliding nuclei. At a later
stage, if the conditions reached in relativistic heavy ion collisions
are adequate, colorless QGP may be formed via fusion of the
tubes and the color would no longer be confined to hadronic or flux
tube dimensions.

This flux tube initial state, if created, immediately begins its
complicated evolution to ordinary hadronic matter --- directly or
passing through the QGP state. The question is, which probes can
provide direct information about this flux tube crossover form of the
matter? Considering the longitudinal flux--tube--like field fluctuations
oriented along the reaction axis which cannot be realized in a pure
hadronic case or in a case of equilibrium QGP, we suggest a new
backgroundless signature of the formation of the initial stage in
contrast to most known experimental phenomena which do not give clear
information about the pre-equilibrium stage.

Let's consider the flux tube stage for the central nucleus--nucleus
collision (Fig.~\ref{fig1}). The tube--like fluctuations of the color
fields are oriented along the collision axis. This anisotropy of the
field fluctuations affects the secondary particle distributions.

For a first estimation in a frame of the classic approach, consider a
hardly scattered parton which is moving in a direction transverse to the
reaction axis. This parton may create a flux tube which may decay to a
hard jet. An interaction of this parton with QCD fields in the
longitudinal flux tubes affects the parton's momentum distribution
which determines the distribution of secondary particles in the jet
relative to the jet axis. In particular, bremsstrahlung gluon radiation
takes place in a tangential direction by the quark crossing the strong
gluon field in each tube. This is the main dynamical reason for an
asymmetry: the mean square projection of momentum of secondary
particles in the jet along the reaction axis (Fig.~\ref{fig1},
axis~$z$) becomes larger than in the transverse direction
(axis~$x$). Because the fluctuations of the fields are isotropic in a
vacuum, hadron matter, or QGP states, the distribution of the jet
particles in these cases is axially symmetrical relative to the jet
axis.

Our estimations show that this asymmetry can be visible in
nucleus--nucleus collisions if the flux tube stage takes place. The
value of the asymmetry can be characterized by the ratio
$R_a=\ptz/\ptx$.  Let's estimate the value of the asymmetry of the
parton distribution relative to axis of its initial direction of
movement (immediately after the hard scattering).

\sloppy
$R_a$ depends on energy of colliding nuclei because of energy
dependence of the mean number of flux tubes and mean square transverse
momentum of particles in a jet relative to jet axis. Consider
$e^+e^-$--annihilation to estimate the energy dependence of
\mbox{$\ptx=\ptz$}. In this case charge particle
multiplicity can be parametrized as \cite{GR1}
\begin{equation}
M_{\rm ch}(W)=a+b\ln W^2+c\ln^2 W^2,
\end{equation}
where $a=3.33,\ b=-0.40,\ c=0.26$ and W is invariant mass
of $e^+e^-$ pair.The total particle multiplicity can be roughly
estimated as $M(W)=\ ^3/_2M_{\rm ch}(W)$, and the value of $\ptx$
can be expressed from sphericity~$S$~\cite{GR1}
\begin{equation} \label{S}
S=\frac{3}{2}\frac{\sum_i(p_{ix}^2+p_{iz}^2)}{\sum_ip_i^2},
\end{equation}
and the asymptotic dependence of $W$~\cite{GR3}
\begin{equation} \label{asimpt}
S\propto0.8 W^{-1/2}.
\end{equation}
If the jet is not too wide ($\ptxs+\ptzs\ll\ptys\simeq{<}p^2{>}$, see
Fig.~\ref{fig1}), then $W\approx M{<}p^2{>}^{1/2}$ and
\begin{equation} \label{spi}
\sum_ip_i^2=M\left(\frac{W}{M}\right)^2=\frac{W^2}{M}.
\end{equation}
From (\ref{spi}) we get
\begin{equation} \label{pth}
\ptx=\sqrt{\frac{1}{2}{<}p_x^2+p_z^2{>}}=\sqrt{\frac{1}{2M}\sum_i
(p_{ix}^2+p_{iz}^2)}\simeq0.52\frac{W^{3/4}}{M(W)}.
\end{equation}
For SpS, RHIC and LHC energies we obtain $\ptx=0.26$, 0.44 and
1.8~GeV$/c$ respectively.

\fussy
Now let's estimate the additional momentum of the hard transverse quark
crossed the longitudinal flux tubes. Consider the simplest case of
static cylindrical flux tubes in ground states with a constant uniform
chromoelectric field pointed along the axis of symmetry (the reaction
axis).

\begin{figure}

\setlength{\unitlength}{1mm} \begin{center}
\begin{picture}(85,51)(-9,0)    \thicklines
\put(40,12.5){\circle*{5}}      \thicklines
\put(15,20){\circle*{5}}        \put(15,30){\circle{5}}
\put(15,40){\circle*{5}}        \put(38.25,12.5){\line(0,-1){11}}
\put(43.5,2){tube}              \put(41.75,12.5){\line(0,-1){11}}
\multiput(15.3,21.75)(0,10){3}{ \begin{picture}(10,10)
\put(0,0){\line(1,0){48.5}}     \put(0,-3.5){\line(1,0){48.5}}
\end{picture}}                  \thinlines
\put(40,20){\vector(1,0){10}}   \put(40,30){\vector(-1,0){10}}
\put(40,40){\vector(1,0){10}}   \thicklines
\put(48,23){$\vec{F}$}          \put(43,9){$q_{1}$}
\put(17,15){$q_{2}$}            \thinlines
\put(7,28){\oval(4,26)[bl]}     \put(3,28){\oval(4,4)[tr]}
\put(7,32){\oval(4,26)[tl]}     \put(3,32){\oval(4,4)[br]}
\put(-2,29){$N$}                \put(40,10){\vector(0,1){38}}
\put(20,14){\oval(20,25)}       \put(12,8){$V$}
\multiput(18,2.5)(1,0){5}{      \line(1,1){5}}
\multiput(18,3)(1,1){5}{        \line(1,0){5}}
\put(29,1.5){$S$}
\end{picture} \end{center}
\caption{
An illustration of the calculation of quark momentum and for discussion
of the interaction between a hard scattered quark and flux tubes.
\label{fig2}}
\end{figure}
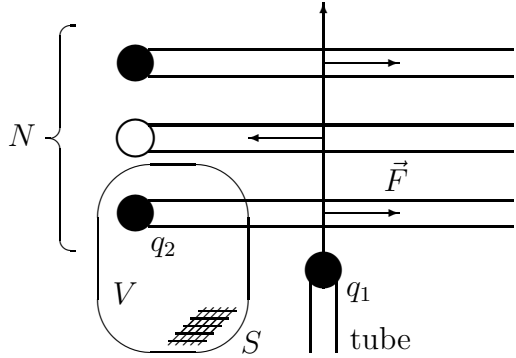

Then consider a quark ($q_{1}$) passing through the flux tube
transverse to the tube axis. Static tube approximation can be used for
the piece of the tube (crossed by the quark) which is almost at rest in
the equal velocity frame. Because in static tubes chromomagnetic fields
vanish, and soft interactions of the hard quark $q_{1}$ with its own
(transverse) tube (Fig.~\ref{fig2}) can be neglected for a study of the
transverse dynamics, the force acting on the quark can be written as
\begin{equation}
{\bf F}=\frac{g}{2}\lambda^{a}_{1}{\bf E}^{a}, \label{eq.1}
\end{equation}
where $g$ is the QCD coupling constant, $\lambda^{a}$ are the Gell-Mann
matrices, the lower index $1$ means that the matrix acts on $q_{1}$ and
${\bf E}^{a}$ is the chromoelectric field. For simplicity, the flux
tube lengths are assumed to be much greater than their radii, and the
flux tube radii to be small enough to neglect the overlap of the tubes.
For the quark induced case the tube ends have triplet color
representations. The quark $q_{1}$ moves with almost light velocity,
therefore, in the thin tube case, the end quarks have no time to change
their color states when $q_{1}$ crosses the flux tube. So the values
of ${\bf E}^{a}$ are distributed among the tubes in accordance with
their color states and can be considered time independent.

Find the change of the mean transverse momentum of the hard quark after
crossing one flux tube with an arbitrary distribution of the gluonic
field over the tube section
\begin{equation}
{\bf E}^{a}={\bf E}^{a}(x,y). \label{eq.2}
\end{equation}
Let axes $x$ and $y$ be the coordinate system in the section where $q_1$
crosses the tube, let axis $y$ be pointed along $q_{1}$'s direction of
motion and let $\xi$ be the distance between axis $y$ and line of $q_1$
movement. Then $q_{1}$ will cross the tube surface in points
$(\xi,y_{1}(\xi))$ and $(\xi,y_{2}(\xi))$.  The quark momentum change
may be written as
\begin{equation}
\triangle{\bf p}(\xi)=\int{\bf F}(\xi,t)\,dt=\int\limits^{y_{2}(\xi)}
_{y_{1}(\xi)}{\bf F}(\xi,y)\,dy=\int\limits^{y_{2}
(\xi)}_{y_{1}(\xi)}g\frac{1}{2}\lambda^{a}_{1}{\bf E}^{a}(\xi,y)\,dy.
\label{eq.3}
\end{equation}
Here and below the light velocity ($c$) is assumed to be $c=1$. Average
$\triangle{\bf p}$ over all $\xi$ ($\rho_{\rm tube}$ is the flux
tube radius):
\begin{eqnarray}
\triangle{\bf p}&=&\frac{1}{\int\limits_{-\rho_{\rm tube}}^{\rho_{
\rm tube}}d\xi}\int\limits_{-\rho_{\rm tube}}^{\rho_{\rm
tube}}d\xi\int\limits^{y_{2}(\xi)}_{y_{1}(\xi)}g\frac{1}{2}\lambda^{a}_
{1}{\bf E}^{a}(\xi,y)\,dy=     \nonumber     \\
&=&\frac{1}{2\rho_{\rm tube}}\int\limits_{S}g\frac{1}{2}\lambda^{a}
_{1}{\bf E}^{a}(x,y)\,dS.     \label{eq.4}
\end{eqnarray}
Here $S$ is a transverse flux tube section. Applying the Gauss theorem
(see Fig.~\ref{fig2}) we get:
\begin{eqnarray}
\triangle{\bf p}&=&\frac{\widehat{\bf z}}{2\rho_{\rm tube}}g\frac
{1}{2}\lambda_{1}^{a}\oint\limits_{S}{\bf E}^{a}\,d{\bf S}=
\frac{\widehat{\bf z}}{2\rho_{\rm tube}}g\frac{1}{2}\lambda_{1}^
{a}\int\limits_{V}\dv{\bf E}^{a}\,dV=     \nonumber     \\
&=&\frac{\widehat{\bf z}}{2\rho_{\rm tube}}\,g\frac{1}{2}\lambda_
{1}^{a}\,g\frac{1}{2}\lambda_{2}^{a}=\widehat{\bf z}\frac{g^{2}}{8\rho_{
\rm tube}}\lambda^{a}_{1}\lambda^{a}_{2},
\label{eq.5}
\end{eqnarray}
where $g\frac{1}{2}\lambda^{a}_{2}$ is the "color charge" of the quark
$q_{2}$ and $\widehat{\bf z}$ is the unit vector in the longitudinal
tube direction. This result does not depend on the field distribution
over the tube section.

For estimation of the asymmetry of distribution of particles in the jet
mentioned above we need to know ${<}\triangle{\bf p}^{2}{>}$:
\begin{equation}
{<}\triangle{\bf
p}^{2}{>}=\frac{g^4}{64\rho_{\rm tube}^2}{<}(\lambda
^a_1\lambda^a_2)^2{>}_c. \label{eq.6}
\end{equation}
Here the index $c$ means averaging over the color states of $q_{1}$ and
$q_{2}$. To calculate ${<}(\lambda^a_1\lambda^a_2)^2{>}_c$ let's
consider matrices $\lambda^a_1$ and $\lambda^a_2$ as vectors in
eight-dimensional space. Choose the axis $x_8$ to be directed along the
vector $\lambda^a_1$ (its distribution is assumed to be isotropic).
Take into account that the mean square of the scalar product
$(\lambda^a_1\lambda^a_2)^2$ averaged over all $\lambda^a_2$ directions
and the mean value of the square of the projection of $\lambda^a_2$
onto the direction of $\lambda^a_1$ (i.e.  ${<}(\lambda^8)^2{>}_c$)
multiplied by the absolute value of $\lambda^a_1$ squared are equal to
each other. This takes place because the color states of $q_{1}$ and
$q_{2}$ are assumed to be independent.  Taking into account that all
mean square components of the vector $\lambda^a$ are equal to each
other we get:
\begin{eqnarray}
{<}\lambda^2{>}_c={<}\lambda^a\lambda^a{>}_c&=&\sum_{a=1}^{8}{<}(\lambda
^a)^2{>}_c=8{<}(\lambda^8)^2{>}_c, \nonumber \\
{<}(\lambda^8)^2{>}_c&=&\frac{1}{8}{<}\lambda^2{>}_c=\frac
{1}{8}\lambda^2. \label{eq.7}
\end{eqnarray}
For $\lambda^2=\frac{16}{3}$:
\begin{equation}
{<}(\lambda^a_1\lambda^a_2)^2{>}_c=\frac{1}{8}\left(\frac{16}
{3}\right)^{\!2}. \label{eq.8}
\end{equation}
Let $\triangle p=\sqrt{{<}\triangle{\bf p}^2{>}_c}$. Then after
crossing one flux tube
\begin{equation}
\triangle
p=\frac{g^2}{8\rho_{\rm tube}}\sqrt{{<}(\lambda^a_1\lambda
^a_2)^2{>}_c}=\frac{4\pi\alpha_s}{3\sqrt{2}\rho_{\rm tube}},
\label{eq.9}
\end{equation}
where $\alpha_s=\frac{g^2}{4\pi}$ is the QCD fine structure constant.

For a tube radius $\approx0.3$~Fm~\cite{2} we get for QCD running
coupling constant
$\alpha_s(\frac{\pi}{\Lambda_{\rm QCD}\rho_{\rm tube}})=0.3$
(in 3--loop approximation) and \mbox{$\triangle p\approx0.6$~GeV}. In
the cases of SpS and RHIC (see Table~\ref{table2} below) this value is
greater than or close to $\ptx$ and asymmetry
\mbox{$R_a\simeq{}$1.7--2.5} is larger than in the case of LHC
($R_a\simeq1.05$).

This asymmetry is expected to be visible for hadron--hadron collisions,
and increases together with the atomic number $A$ of the colliding
nuclei because the quark crosses more and more tubes. In this paper we
consider the jet asymmetry in the large $A$ case for the reaction of
$^{238}U{+}^{238}U$.

Now let's estimate the average number of flux tubes crossed by the quark
in its traveling across the reaction zone in central nucleus--nucleus
collisions. For kinematical reasons in addition to the quark~$q_1$
there is a quark $q_1'$ hardly collided with $q_{1}$ and moving in the
opposite direction from $q_{1}$. The mean distance travelled by each
quark in the nuclear reaction zone with uniform distribution of the
flux tubes is then
\begin{equation}
{<}l{>}=\frac{8R}{3\pi}. \label{eq.10}
\end{equation}
Here $R$ is the reaction zone radius (in the case of the central
interactions of nuclei it is almost equal to the radius of the lightest
nucleus).

We assume that the mean number of flux tubes is about $3/2$ for each
nucleon because for each nucleon there are four possible numbers
$0,1,2,3$ of the tubes encountered with almost equal probability.
This rough estimation is in good agreement with FRITIOF \cite{fr}
calculations.

Consider a central collision of two identical nuclei with atomic
number~$A$. In this case the number of flux tubes per unit area of the
transverse section is
\begin{equation}
n=\frac{\frac{3}{2}A}{\pi R^2}=\frac{3A^{1/3}}{2\pi r^2_0}.
\label{eq.11}
\end{equation}
Here $R=r_0A^{1/3}$ is the nucleus radius. Then the average number of
flux tubes crossed by $q_{1}$ will be
\begin{equation}
N=2\rho_{\rm tube}{<}l{>}n=\frac{8A^{2/3}}{\pi^2r_0}\rho_{\rm
tube}. \label{eq.12}
\end{equation}
For $^{238}U$ we get $N\approx8$.

Let's find the average transverse momentum of quark $q_{1}$ after it
crosses $N$ flux tubes. After crossing $m$ flux tubes
\begin{eqnarray}
{<}\triangle{P_z}^2_m{>}&=&{<}(\triangle{P_z}_{m-1}+\triangle{p_z})^2
{>}= \nonumber   \\   &=&{<}\triangle{P_z}^2_{m-1}{>}+{<}\triangle{p_z}
^2{>}+2{<}\triangle{P_z}_{m-1} \triangle{p_z}{>}.
\label{eq.13}
\end{eqnarray}
Here $\triangle{P_z}_{m-1}$ is the transverse momentum of $q_{1}$
after it crossed $m-1$ flux tubes. The last term vanishes because
$\triangle{P_z}_{m-1}$ and $\triangle{p_z}$ may be pointed in any
direction along the tube axis independently. So, we get
\begin{equation}
{<}\triangle{P_z}^2_m{>}={<}\triangle{P_z}^2_{m-1}{>}+{<}\triangle{p_z}
^2{>}. \label{eq.14}
\end{equation}
And
\begin{eqnarray}
{<}\triangle P{>}&=&\sqrt{{<}\triangle{P_z}^2_N{>}}\simeq\sqrt{N(
\triangle p)^2}=\sqrt{N}\triangle p \label{eq.15a} \\
{<}\triangle P{>}&=&\sqrt{\frac{8A^{2/3}}{\pi^2r_0}\rho_{\rm tube}}
\frac{4
\pi\alpha_s(\rho_{\rm tube})}{3\sqrt{2}\rho_{\rm tube}}=
\frac{2\pi
A^{1/3}}{\sqrt{6r_0}}\frac{\alpha_s(\rho_{\rm tube})}
{\sqrt{\rho_{\rm tube}}}\approx \nonumber \\
&\approx&1.7\,\,{\rm GeV}/c. \label{eq.15b}
\end{eqnarray}

This estimation is accurate for the case of non-overlapping flux tubes.
If the flux tube radii are not too large then this assumption is quite
reasonable.

Let's estimate the asymmetry of particle distribution in a jet
characterized by $R_a=\ptz/\ptx$. The value of $\ptz$ is determined by
average transverse momentum of the hard quark ${<}\triangle P{>}$ and by
typical transverse momentum of particles in the jet~$\ptx$:
\begin{eqnarray}
R_a&=&\frac{\ptz}{\ptx}=\frac{\sqrt{{<}\triangle P{>}^2+\ptxs}}{\ptx}=
\sqrt{1+\frac{{<}\triangle
P{>}^2}{\ptxs}}= \nonumber \\
&=&\sqrt{1+\frac{16}{3}\frac{A^{2/3}\alpha_s^2}{r_0\rho_{
\rm tube}\ptxs}}.
\end{eqnarray}
The values of asymmetry $R_a$ at various energies of colliding nuclei
are shown in Table~\ref{table2} (see below). One can see that in the
cases of SpS and RHIC $R_a\gg1$ and so large asymmetry can be measured
experimentally. In the case of LHC the asymmetry is not very large but
may be observed in high statistics experiments.

It should be noted that our calculation of quark interaction with flux
tubes is not complete yet. For the first step we concentrated on
the subject of phenomenon and did not take into account the quark's
interaction with fields in its own flux tube or the interactions
between tubes. We consider only quark jets here although we would
expect the more noticeable effect for gluon-induced jets. The flux
tubes are assumed to be non-overlapping. In Table~\ref{table1} we
present the results of our estimation (based on event generator
FRITIOF) of the mean numbers of essentially overlapped tubes for
different tube radii for the reaction of $^{238}U+\,^{238}U$. We
suppose here that the tubes are essentially overlapped when the
distance between their axes is less than the tube radius. The resulting
value~(\ref{eq.15b}) takes into account only the most important
contributions and needs in more precise calculations which we plan to
make in our following works. The small value of tube radius
0.2--0.3~Fm are of great interest (in connection with
percolation~\cite{Per} and lattice calculations~\cite{Lat}). One can
see from the Table~\ref{table1} that in this case the most of tubes are
not overlapped.

\begin{table}
\caption{
Mean numbers of $n-$fold overlapped flux tubes for different tube radii
($n=0$ corresponds to the not overlapped single flux tube).
\label{table1}}
\begin{center}
\begin{tabular}{|c|c|c|c|c|} \hline
& \multicolumn{4}{c|}{Tube radius (Fm)} \\ \cline{2-5}
n & 0.1   & 0.2   & 0.3   & 0.4   \\ \hline \hline
0 & 330.9 & 261.6 & 175.9 & 102.7 \\
1 & 25.4  & 82.0  & 123.9 & 120.9 \\
2 & 0.7   & 11.6  & 43.5  & 81.3  \\
3 & 0.0   & 1.6   & 10.5  & 35.3  \\
4 & 0.0   & 0.2   & 2.3   & 12.0  \\ \hline
\end{tabular}
\end{center}
\end{table}

It should be noted that a jet's properties can be studied in the
experiment if its angular size is sufficiently large to cover more than
one detector module and if there is no dominance of essentially
overlapped jets.

\begin{figure}

\setlength{\unitlength}{1mm} \begin{center}
\begin{picture}(85,60)(-5,2)    \thicklines
\put(6,37){\circle{12}}         \put(64,37){\circle{12}}

\put(6,40){\circle*{3}}         \put(64,40){\circle*{3}}
\put(3.5,35){\circle*{3}}       \put(8.5,35){\circle*{3}}
\put(61.5,35){\circle*{3}}      \put(66.5,35){\circle*{3}}

\put(6,40){\vector(1,0){15}}    \put(21,40){\line(1,0){8}}
\put(29,46){\oval(12,12)[br]}   \put(35,46){\vector(0,1){12}}
\put(18,43){$x_{1}p$}           \put(47,43){$-x_{2}p$}
\put(64,40){\vector(-1,0){12}}  \put(52,40){\line(-1,0){11}}
\put(41,34){\oval(12,12)[tl]}   \put(35,34){\vector(0,-1){28}}

\put(0,27){\line(1,0){10}}      \put(0,25){\line(1,0){10}}
\put(8,28){\line(2,-1){4}}      \put(8,24){\line(2,1){4}}
\put(2,20){$p_0$}
\put(70,27){\line(-1,0){10}}    \put(70,25){\line(-1,0){10}}
\put(62,28){\line(-2,-1){4}}    \put(62,24){\line(-2,1){4}}
\put(62,20){$-p_0$}             \put(35,6){\vector(1,0){7}}

\thinlines
\put(35,34){\vector(-1,-4){7}}  \put(35,34){\vector(1,-4){7}}
\put(-3,37){\vector(1,0){77}}   \put(72,34){$z$}
\put(35,11.33){\line(-5,-1){28.67}}
\put(0,0.33){$\pty$}            \put(35,0.33){$\ptx$}
\thicklines                     
\put(40,50){$(x_{1}=x_{2}=1/3)$}

\end{picture} \end{center}

\caption{
An illustration of the angle width of the jet.
\label{fig3}}
\end{figure}
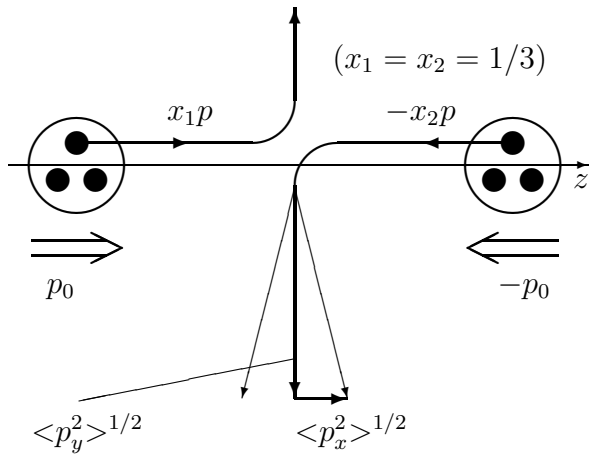

To estimate the angular size of the jets let's consider a jet in the
equal velocity frame where the partons have been elastically scattered
off each other at the angle of 90 degrees. Each parton carries about
$x=1/3$ of the total nucleon momentum $p_0$ (sea quarks and gluons do
not change strongly this value). The angular size of a jet is
determined by (Fig.~\ref{fig3})
\begin{eqnarray}
\sin\frac{\theta}{2}=\frac{\ptx}{\pty}\simeq M\frac{\ptx}{2p_0/3}=
\frac{M(W)}{W}\ptx,
\end{eqnarray}
where $\ptx$ is shown in (\ref{pth}) and the factor 2 in the denominator
appears because $M$ is the multiplicity of the two jets propagating in
opposite directions. In Table~\ref{table2} we present values of $p_0$,
$\ptx$, $\ptz$, $R_a$ and $\theta$ for SpS, RHIC and LHC energies. One
can see that in all cases the angular size of the jets is large enough
to be detected experimentally.

\begin{table}
\caption{
Nucleon momentum~$p_0$, invariant mass of hardly scattered partons $W$,
particle multiplicity of di-jet~$M$, typical scale of the particle
transverse momenta in jets in the beam axis direction~($\ptz$) and in
direction perpendicular to the jet and beam axes ($\ptx$), value of jet
asymmetry $R_a$ and jet angular size $\theta$ for different colliders.
\label{table2}}
\begin{center}
\begin{tabular}{|c|c|c|c|} \hline
Collider & SPS & RHIC & LHC \\ \hline \hline
$p_0$, GeV$/c$ & 10 & 100 & 3150 \\ \hline
$W$, GeV & 20/3 & 200/3 & 6300/3 \\ \hline
$M$ & 8.33 & 27.5 & 87.1 \\ \hline
$\ptx$, GeV$/c$ & 0.26 & 0.44 & 1.8 \\ \hline
$\ptz$, Gev$/c$ & 1.7 & 1.7 & 2.5 \\ \hline
$R_a$ & 6.7 & 4.0 & 1.4 \\ \hline
$\theta,^\circ$ & 38 & 21 & 8.7 \\ \hline
\end{tabular}
\end{center}
\end{table}

We have yet no estimations of the numbers of overlapping jets at
different transverse momenta. But in any case there are transverse jets
with high momentum which are rare enough to be single.

Other corrections may arise from the decay of the longitudinal tubes on
fragments due to their rotation in evolution of the flux-tube stage.
The largest asymmetry takes place in the case when the quark crosses
the flux tubes transversely. Because the angular velocity of tube
rotation is greater in case of short tube fragments (the mean square
transverse momentum of fragment ends does not depend on the fragment
lengths), events in which the long tube fragments dominate are more
favorable for studying the asymmetry. In particular such events may be
selected by using the fact that longer tubes accumulate more kinetic
energy in their tension and their fragments after decay should be
concentrated in the middle rapidity range. We plan to present more
detailed calculations of this effect in our future works. In any case
the large value of asymmetry ($\sim670$\%) means that this asymmetry
can be observed in experiments even if the corrections would be taken
into account.

Another reason for decreasing of the asymmetry is the percolation
effect. As shown in~\cite{Per} even in central Pb--Pb collisions string
density is so high that practically all tubes are fused. It means that
there is no jet asymmetry. But it is not yet known if this picture
is real or not. Observing any asymmetry or its absence in
experiments may give information about the way of evolution of heavy
ion collisions.

In conclusion it should be noted that a new possible manifestation of
the formation of the initial nonequilibrium flux tube stage for the QGP
transition in high energy heavy ion collisions is suggested: the
longitudinal anisotropy in the flux tube fluctuations of color electric
fields formed in the pre-equilibrium (flux tube) stage leads to a
strong azimuthal asymmetry in the particle distributions in the
hard-transverse jets around their axes. It is concerned with the large
difference between the transverse and longitudinal (with respect to the
beam axis) momentum components of the hardly scattered partons. This
increasing of the longitudinal momentum component takes place due to
the interaction of the transverse jet quark with gluon fields in the
longitudinally oriented flux tubes. The value of the additional
longitudinal momentum of the jet quark after crossing the flux tubes
(${<}\triangle P{>}\approx1.7$~GeV) is sufficiently large
to be observable in experiments. Because this asymmetry is produced by
the axially symmetrical field fluctuations, it should be absent in
both the cases of hadronic matter and QGP and may serve as a
backgroundless signature of the flux tube stage in heavy ion
collisions.

\end{document}